\documentclass[11pt]{article}
\usepackage{bbm}
\usepackage{graphicx}
\usepackage{amsmath}
\usepackage{amssymb}
\usepackage{caption2}
\begin{document}
\bibliographystyle{prsty}
\begin{center}
{\large {\bf \sc{ Analysis of the vertexes $\Omega_Q^*\Omega_Q \phi$
and radiative decays
$\Omega_Q^*\to \Omega_Q \gamma$ }}} \\[2mm]
Zhi-Gang Wang \footnote{E-mail,wangzgyiti@yahoo.com.cn. }    \\
 Department of Physics, North China Electric Power University, Baoding 071003, P. R. China \\
\end{center}

\begin{abstract}
In this article, we study the vertexes  $\Omega_Q^*\Omega_Q \phi$
with the light-cone QCD sum rules, then assume  the vector meson
dominance of the intermediate $\phi(1020)$, and calculate  the
radiative decays $\Omega_Q^*\to \Omega_Q \gamma$.

\end{abstract}


\section{Introduction}

In 2006, the BaBar collaboration reported the first observation of
the ${3\over 2}^+$ baryon $\Omega_{c}^{*}$  in
 the radiative decay $\Omega_{c}^{*}\rightarrow\Omega_c\gamma$, where the
${1\over 2}^+$ baryon $\Omega_c$ was reconstructed in decays to the
final states $\Omega^-\pi^+$, $\Omega^-\pi^+\pi^0$,
$\Omega^-\pi^+\pi^-\pi^+$ and $\Xi^- K^-\pi^+\pi^+$ \cite{OmegaC}.
The $\Omega_{c}^{*}$ lies about $70.8 \pm 1.0 \pm 1.1 \rm{MeV}$
above the $\Omega_c$, and  is the last singly-charm baryon  with
zero orbital momentum observed experimentally \cite{Rosner95}. In
2008, the D0 collaboration reported the first observation of the
doubly strange baryon $\Omega_{b}^{-}$ in the decay channel
$\Omega_b^- \to J/\psi\thinspace\Omega^-$ (with
$J/\psi\to\mu^+\mu^-$ and $\Omega^-\to\Lambda K^-\to p\pi^- K^-$) in
$p\bar{p}$ collisions at $\sqrt{s}=1.96$ TeV \cite{OmegabD0}. The
experimental value $M_{\Omega_b^-}=6.165\pm 0.010\thinspace \pm
0.013\thinspace \, \rm{GeV}$ is about $0.1 \, \rm{GeV}$ larger than
the existing theoretical calculations (see Ref. \cite{Wang0809Omega}
for a short review on the relevant literatures); however,
 the CDF collaboration did not confirm the   measured
 mass \cite{OmegabCDF}, i.e. they  observed the mass of the $\Omega^-_b$   is about $6.0544\pm 0.0068
\pm 0.0009 \,\rm{GeV} $, which is consistent with the existing
theoretical calculations.

  By now, the ${1\over 2}^+$
antitriplet states ($\Lambda_c^+$, $\Xi_c^+,\Xi_c^0)$,  and the
${1\over 2}^+$ and ${3\over 2}^+$ sextet states
($\Omega_c,\Sigma_c,\Xi'_c$) and ($\Omega_c^*,\Sigma_c^*,\Xi'^*_c$)
have been well established; while the corresponding  $S$-wave
bottom baryons are far from complete, only the $\Lambda_b$,
$\Sigma_b$, $\Sigma_b^*$, $\Xi_b$, $\Omega_b$ have been observed
\cite{PDG}. Those heavy baryons  are particularly interesting for
studying dynamics of the light quarks in the presence  of a heavy
quark. In the heavy quark limit, the three light  quarks form an
$SU(3)$ flavor triplet, ${\bf 3}\times {\bf 3}={\bf \bar 3}+{\bf
6}$, two light quarks can form diquarks of a symmetric sextet  and
an antisymmetric antitriplet \cite{ReviewH1,ReviewH2}.

The light-cone QCD sum rules are a powerful theoretical tool in
studying the ground state heavy baryons, they carry out the operator
product expansion near the light-cone $x^2\approx 0$ instead of the
short distance $x\approx 0$, while the nonperturbative hadronic
matrix elements  are parameterized by the light-cone distribution
amplitudes   instead of the vacuum condensates
\cite{LCSR89,LCSR,LCSRreview}. The nonperturbative
 parameters in the light-cone distribution amplitudes are calculated with  the conventional QCD  sum rules
 and the  values are universal.

In Ref.\cite{WangVMD}, we assume the charm  mesons $D_{s0}(2317)$
and  $D_{s1}(2460)$ with the spin-parity $0^+$ and $1^+$
respectively are
  the conventional $c\bar{s}$
states, and  calculate the strong coupling constants $\langle D_s^*
\phi | D_{s0} \rangle$ and$\langle D_s \phi | D_{s1} \rangle$  with
the light-cone QCD sum rules, then take the vector meson dominance
of the intermediate $\phi(1020)$,  study the radiative decays
$D_{s0}\to D_s^* \gamma $ and $D_{s1}\to D_s \gamma $. In previous
works \cite{Wang0809Omega,Wang0704} (see also
Refs.\cite{Huang08,Narison09}), we  have calculated  the masses and
the pole residues of the $\frac{1}{2}^+$ heavy baryons  $\Omega_Q$
and the $\frac{3}{2}^+$ heavy baryons  $\Omega_Q^*$   with the QCD
sum rules.  In this article, we extend our previous works to  study
the vertexes $\Omega_Q^*\Omega_Q \phi$ with the light-cone QCD sum
rules \footnote{The results of the strong coupling constants among
the nonet vector mesons, the octet baryons and  the decuplet baryons
will be presented  elsewhere.}, then assume  the vector meson
dominance of the intermediate $\phi(1020)$, and calculate the
radiative decays $\Omega_Q^*\to \Omega_Q \gamma$. In
Ref.\cite{Aliev0901}, Aliev et al study  the radiative  decays
$\Sigma_Q^*\to\Sigma_Q\gamma$, $\Xi_Q^*\to\Xi_Q\gamma$ and
$\Sigma_Q^*\to\Lambda_Q\gamma$
 with the light cone
QCD sum rules.

There have been many works dealing with the strong coupling
constants of the pseudoscalar (scalar) octet mesons  and vector
nonet mesons with the  baryons. The $\rho NN$, $\rho\Sigma\Sigma$,
$\rho\Xi\Xi$ and other strong coupling constants of the nonet vector
mesons with the octet baryons have been calculated using the light
cone QCD sum rules \cite{Zhu99,Wang0701,Aliev0905}. In
Refs.\cite{Aliev2006,Aliev0908}, Aliev et al study the strong
coupling constants of the pseudoscalar octet mesons with the octet
(and decuplet) baryons   comprehensively. In
Refs.\cite{Wang0707,Wang0809}, we study the strong decays
$\Delta^{++} \to p \pi$, $\Sigma^*\to \Sigma \pi$ and  $\Sigma^*\to
\Lambda \pi$ using the light-cone QCD sum rules. Moreover, the
coupling constants of the vector mesons $\rho$ and $\omega$ with the
baryons are studied with the external field QCD sum rules
\cite{Erkol2006}.

The article is arranged as: in Section 2, we derive the strong
coupling constants  $g_1$, $g_2$ and $g_3$ of the vertexes
$\Omega_Q^*\Omega_Q \phi$ with the light-cone QCD sum rules; in
Section 3, the numerical result and discussion; and Section 4 is
reserved for conclusion.
\section{ The vertexes  $\Omega_Q^*\Omega_Q \phi$ with light-cone QCD sum rules}
We parameterize the vertexes $\Omega_Q^*\Omega_Q \phi$ with three
tensor structures  due to  Lorentz invariance and introduce three
strong  coupling constants $g_1$, $g_2$ and $g_3$ \cite{ANN1973},
\begin{eqnarray}
\langle \Omega_Q(p+q)|\Omega_Q^*(p)\phi(q)
\rangle&=&\overline{U}(p+q) \left[
g_1(q_\mu\not\!\!{\epsilon}-\epsilon_\mu \not\!\!{q})\gamma_5+g_2
(P\cdot\epsilon q_\mu-P\cdot q \epsilon_\mu) \gamma_5
\right.\nonumber\\
&&\left. +g_3(q\cdot \epsilon q_\mu-q^2
\epsilon_\mu)\gamma_5\right]U^\mu(p)
\nonumber \\
&=&\epsilon_\mu\overline{U}(p+q) \Gamma^{\mu\nu}U_\nu(p) \, ,
\end{eqnarray}
where the $U(p)$ and $U_\mu(p)$ are the Dirac spinors of the heavy
baryons $\Omega_Q$ and $\Omega_Q^*$ respectively,   the
$\epsilon_\mu$ is the polarization vector of the meson $\phi(1020)$,
and $P=\frac{2p+q}{2}$.

In the following, we write down the
 two-point correlation function $\Pi_\mu(p,q)$,
\begin{eqnarray}
\Pi_\mu(p,q)&=&i \int d^4x \, e^{-i p \cdot x} \,
\langle 0 |T\left\{ J(0)\bar{J}_\mu(x)\right\}|\phi(q)\rangle \, , \\
J(x)&=& \epsilon^{ijk}  s^T_i(x)C\gamma_\mu s_j(x) \gamma_5 \gamma^\mu Q_k(x)  \, ,  \nonumber \\
J_\mu(x)&=& \epsilon^{ijk}  s^T_i(x)C\gamma_\mu s_j(x)  Q_k(x)\, ,
\end{eqnarray}
where $Q=c$ and $b$, the $i,j,k$ are color indexes,  the Ioffe type
heavy baryon currents $J(x)$ and $J_\mu(x)$
 interpolate the $\frac{1}{2}^+$ baryons  $\Omega_Q$ and the  $\frac{3}{2}^+$
 baryons  $\Omega_Q^*$  respectively \cite{Wang0809Omega,Wang0704}, the external  vector state $\phi(1020)$ has the
four momentum $q_\mu$ with $q^2=M_\phi^2$ .

Basing on the quark-hadron duality \cite{SVZ79,PRT85}, we can insert
a complete set  of intermediate hadronic states with the same
quantum numbers as the current operators $J(x)$ and $J_\mu(x)$ into
the correlation function $\Pi_{\mu}(p,q)$  to obtain the hadronic
representation. After isolating the ground state contributions from
the pole terms of the baryons $\Omega_Q$ and $\Omega_Q^*$, we get
the following result,
\begin{eqnarray}
\Pi_{\mu }(p,q)&=&\frac{\langle0| J(0)| \Omega_Q(q+p)\rangle\langle
\Omega_Q(q+p)| \Omega_Q^*(p) \phi(q) \rangle \langle
\Omega_Q^*(p)|\bar{J}_\mu(0)| 0\rangle}
  {\left[M_{\Omega_Q}^2-(q+p)^2\right]\left[M_{\Omega_Q^*}^2-p^2\right]}  + \cdots \nonumber \\
&=& \frac{\lambda_{\Omega_Q}\lambda_{\Omega_Q^*}}
{\left[M_{\Omega_Q}^2-(q+p)^2\right]\left[M_{\Omega_Q^*}^2-p^2\right]}
\left\{g_1 \left[M_{\Omega_Q}+M_{\Omega_Q^*}
\right]\!\not\!{\epsilon}\!\not\!{p} \gamma_5 q_\mu
    \right.\nonumber\\
    && -g_1\left[M_{\Omega_Q}+M_{\Omega_Q^*}
\right]\!\not\!{q}\!\not\!{p} \gamma_5\epsilon_\mu
+g_2\!\not\!{q}\!\not\!{p} \gamma_5 p\cdot\epsilon
q_\mu-g_2\!\not\!{q}\!\not\!{p} \gamma_5 q \cdot p \epsilon_\mu
\nonumber\\
&&\left.-\frac{g_2}{2}\!\not\!{q}\!\not\!{p} \gamma_5 q^2
\epsilon_\mu-g_3 \!\not\!{q}\!\not\!{p} \gamma_5 q^2
\epsilon_\mu+\cdots\right\}+\cdots \, ,
\end{eqnarray}
where the following definitions have been used,
\begin{eqnarray}
\langle 0| J (0)|\Omega_Q(p)\rangle &=&\lambda_{\Omega_Q} U(p,s) \, , \nonumber \\
\langle 0| J_\mu (0)|\Omega_Q^*(p)\rangle &=&\lambda_{\Omega_Q^*} U_\mu(p,s) \, , \nonumber \\
\sum_sU(p,s)\overline {U}(p,s)&=&\!\not\!{p}+M_{\Omega_Q} \, , \nonumber \\
\sum_s U_\mu(p,s) \overline{U}_\nu(p,s)
&=&-(\!\not\!{p}+M_{\Omega_Q^*})\left( g_{\mu\nu}-\frac{\gamma_\mu
\gamma_\nu}{3}-\frac{2p_\mu p_\nu}{3M_{\Omega_Q^*}^2}+\frac{p_\mu
\gamma_\nu-p_\nu \gamma_\mu}{3M_{\Omega_Q^*}} \right) \,  .
\end{eqnarray}

 The current $J_\mu(x)$ couples
not only to the  spin-parity $J^P=\frac{3}{2}^+$ states, but also to
the   spin-parity $J^P=\frac{1}{2}^-$ states.
 For a generic $\frac{1}{2}^-$ resonance   $\widetilde{\Omega}_Q^*$ ,
\begin{eqnarray}
 \langle0|J_{\mu}(0)|\widetilde{\Omega}_Q^*(p)\rangle=\lambda_{*}
 (\gamma_{\mu}-4\frac{p_{\mu}}{M_{*}})U^{*}(p,s) \, ,
\end{eqnarray}
where $\lambda^{*}$ is  the  pole residue  and $M_{*}$ is the mass.
The spinor $U^*(p,s)$  satisfies the usual Dirac equation
$(\not\!\!p-M_{*})U^{*}(p)=0$.
 If we choose the tensor structures $\!\not\!{\epsilon}\!\not\!{p} \gamma_5 q_\mu$,
$\!\not\!{q}\!\not\!{p} \gamma_5 p\cdot\epsilon q_\mu$,
$\!\not\!{q}\!\not\!{p} \gamma_5  \epsilon_\mu$, the baryon
$\widetilde{\Omega}_Q^*$  has no contamination.

In the following, we briefly outline the  operator product expansion
for the correlation function  $\Pi_{\mu }(p,q)$  in perturbative QCD
theory. The calculations are performed at the large space-like
momentum regions $(q+p)^2\ll 0$  and $p^2\ll 0$, which correspond to
the small light-cone distance $x^2\approx 0$ required by the
validity of the operator product expansion approach. We write down
the "full" propagator of a massive    quark in the presence of the
quark and gluon condensates firstly \cite{LCSR89,PRT85},
\begin{eqnarray}
S_{ij}(x)&=& \frac{i\delta_{ij}\!\not\!{x}}{ 2\pi^2x^4}
-\frac{\delta_{ij}m_s}{4\pi^2x^2}-\frac{\delta_{ij}}{12}\langle
\bar{s}s\rangle +\frac{i\delta_{ij}}{48}m_s
\langle\bar{s}s\rangle\!\not\!{x}-\frac{\delta_{ij}x^2}{192}\langle \bar{s}g_s\sigma Gs\rangle\nonumber\\
&& +\frac{i\delta_{ij}x^2}{1152}m_s\langle \bar{s}g_s\sigma
 Gs\rangle \!\not\!{x}-\frac{i}{16\pi^2x^2} \int_0^1 dv G^{ij}_{\mu\nu}(vx) \left[(1-v)\!\not\!{x}
\sigma^{\mu\nu}+v\sigma^{\mu\nu} \!\not\!{x}\right]  +\cdots \, ,\nonumber\\
S_Q^{ij}(x)&=&\frac{i}{(2\pi)^4}\int d^4k e^{-ik \cdot x} \left\{
\frac{\delta_{ij}}{\!\not\!{k}-m_Q}
-\frac{g_sG^{\alpha\beta}_{ij}}{4}\frac{\sigma_{\alpha\beta}(\!\not\!{k}+m_Q)+(\!\not\!{k}+m_Q)
\sigma_{\alpha\beta}}{(k^2-m_Q^2)^2}\right.\nonumber\\
&&\left.+\frac{\pi^2}{3} \langle \frac{\alpha_sGG}{\pi}\rangle
\delta_{ij}m_Q \frac{k^2+m_Q\!\not\!{k}}{(k^2-m_Q^2)^4}
+\cdots\right\} \, ,
\end{eqnarray}
where $\langle \bar{s}g_s\sigma Gs\rangle=\langle
\bar{s}g_s\sigma_{\alpha\beta} G^{\alpha\beta}s\rangle$  and
$\langle \frac{\alpha_sGG}{\pi}\rangle=\langle
\frac{\alpha_sG_{\alpha\beta}G^{\alpha\beta}}{\pi}\rangle$, then
contract the quark fields in the correlation function $\Pi_\mu(p,q)$
with Wick theorem, and obtain the result:
\begin{eqnarray}
\Pi_\mu(p,q)&=&2i\epsilon^{ijk}\epsilon^{i'j'k'} \int d^4x e^{-i p
\cdot x} \nonumber \\
&&\left\{\gamma_5\gamma^\alpha S_Q^{kk'}(-x) Tr\left[\gamma_\alpha
\langle 0|s_j(0)\bar{s}_{j'}(x) |\phi(q)\rangle \gamma_\mu
CS_{ii'}^T(-x)C\right]
 \right.\nonumber \\
 &&\left. +\gamma_5\gamma^\alpha S_Q^{kk'}(-x)Tr\left[ \gamma_\alpha S_{jj'}(-x)\gamma_\mu C\langle 0|s_i(0)\bar{s}_{i'}(x) |\phi(q)\rangle^T C\right]\right\}\,  .
\end{eqnarray}
Perform the following Fierz re-ordering to extract the contributions
from the two-particle and three-particle $\phi$-meson light-cone
distribution amplitudes respectively,
\begin{eqnarray}
s^a_\alpha(0) \bar{s}^b_\beta(x)&=&-\frac{1}{12}
\delta_{ab}\delta_{\alpha\beta}\bar{s}(x)s(0)
-\frac{1}{12}\delta_{ab}(\gamma^\mu)_{\alpha\beta}\bar{s}(x)\gamma_\mu
s(0) \nonumber\\
&&-\frac{1}{24}\delta_{ab}(\sigma^{\mu\nu})_{\alpha\beta}\bar{s}(x)\sigma_{\mu\nu}s(0) \nonumber\\
&&+\frac{1}{12}\delta_{ab}(\gamma^\mu
\gamma_5)_{\alpha\beta}\bar{s}(x)\gamma_\mu \gamma_5 s(0)\nonumber\\
&&+\frac{1}{12}\delta_{ab}(i \gamma_5)_{\alpha\beta}\bar{s}(x)i
\gamma_5 s(0) \, , \\
s^a_\alpha(0)
\bar{s}^b_\beta(x)G^{ba}_{\lambda\tau}(vx)&=&-\frac{1}{4}
 \delta_{\alpha\beta}\bar{s}(x)G_{\lambda\tau}(vx)s(0)
-\frac{1}{4} (\gamma^\mu)_{\alpha\beta}\bar{s}(x)\gamma_\mu
G_{\lambda\tau}(vx)
s(0) \nonumber\\
&&-\frac{1}{8} (\sigma^{\mu\nu})_{\alpha\beta}\bar{s}(x)\sigma_{\mu\nu}G_{\lambda\tau}(vx)s(0) \nonumber\\
&&+\frac{1}{4} (\gamma^\mu
\gamma_5)_{\alpha\beta}\bar{s}(x)\gamma_\mu \gamma_5 G_{\lambda\tau}(vx)s(0)\nonumber\\
&&+\frac{1}{4} (i \gamma_5)_{\alpha\beta}\bar{s}(x)i
\gamma_5G_{\lambda\tau}(vx) s(0) \, ,
\end{eqnarray}
 and replace  the hadronic matrix elements (such as the $ \langle
0| {\bar s} (x) \gamma_\mu \gamma_5 s(0) |\phi(q)\rangle$,   etc.)
with
 the corresponding $\phi$-meson light-cone distribution amplitudes, then
  substitute the full $s$  and $Q$ quark propagators into above
correlation function and complete  the integral in the  coordinate
space, finally  integrate over the variable $k$, we can obtain the
correlation function $\Pi_\mu(p,q)$ at the level of quark-gluon
degree of freedom.
 In
calculation, the two-particle and three-particle $\phi$-meson
light-cone distribution amplitudes have been used
\cite{VMLC981,VMLC982,VMLC2003,VMLC2007}, the explicit definitions
are given in the appendix. The parameters in the light-cone
distribution amplitudes are scale dependent and are estimated with
the QCD sum rules \cite{VMLC2003,VMLC2007}. In this article, the
energy scale $\mu$ is chosen to be $\mu=1\,\rm{GeV}$.

Taking double Borel transform  with respect to the variables
$Q_1^2=-p^2$ and $Q_2^2=-(p+q)^2$ respectively,  then subtract the
contributions from the high resonances and continuum states by
introducing  the threshold parameter $s_0$ (i.e. $ M^{2n}\rightarrow
\frac{1}{\Gamma[n]}\int_0^{s_0} ds s^{n-1}e^{-\frac{s}{M^2}}$),
finally we obtain six  sum rules  for the strong coupling constants
$g_{1}$, $g_{2}$ and
$\rm{G}3=-\left(M_{\Omega_Q}+M_{\Omega_Q}^*\right)g_1$ $-M_\phi^2
\left(\frac{g_2}{2}+g_3\right)$ respectively, the explicit
expressions are presented in the appendix.

\section{Numerical result and discussion}
The input parameters are taken as $M_\phi=1.019455\,\rm{GeV}$,
$M_{\Omega_c}=2.6952\,\rm{GeV}$, $M_{\Omega_c^*}=2.7659\,\rm{GeV}$,
$M_{\Omega_b}=6.165\,\rm{GeV}$ \cite{PDG},
$M_{\Omega_b^*}=6.06\,\rm{GeV}$,
$\lambda_{\Omega_c}=(0.075\pm0.01)\,\rm{GeV}^3$,
$\lambda_{\Omega_c^*}=(0.05\pm0.01)\,\rm{GeV}^3$,
$\lambda_{\Omega_b}=(0.10\pm0.01)\,\rm{GeV}^3$,
$\lambda_{\Omega_b^*}=(0.06\pm0.01)\,\rm{GeV}^3$
\cite{Wang0809Omega,Wang0704}, $f_\phi=(0.215\pm0.005)\,\rm{GeV}$,
$f_\phi^{\perp}=(0.186\pm0.009)\,\rm{GeV}$, $a_1^{\parallel}=0.0$,
$a_1^{\perp}=0.0$, $a_2^{\parallel}=0.18\pm0.08$,
$a_2^{\perp}=0.14\pm0.07$, $\zeta^{\parallel}_3=0.024\pm 0.008$,
$\widetilde{\lambda}_3^{\parallel}=0.0$,
$\widetilde{\omega}_3^{\parallel}=-0.045\pm 0.015$,
$\kappa_3^{\parallel}=0.0$, $\omega_3^\parallel=0.09\pm0.03$,
$\lambda_3^\parallel=0.0$, $\kappa_3^\perp=0.0$,
$\omega_3^\perp=0.20\pm0.08$, $\lambda_3^\perp=0.0$,
$\varsigma_4^\parallel=0.00\pm 0.02$,
$\widetilde{\omega}_4^\parallel=-0.02\pm0.01$,
$\varsigma_4^\perp=-0.01\pm 0.03$,
$\widetilde{\varsigma}_4^\perp=-0.03\pm 0.04$,
$\kappa_4^\parallel=0.0$, $\kappa_4^\perp=0.0$
\cite{VMLC2003,VMLC2007}, $m_s=(140\pm 10 )\,\rm{MeV}$,
$m_c=(1.35\pm 0.10)\,\rm{GeV}$, $m_b=(4.7\pm 0.1)\,\rm{GeV}$
\cite{PDG},  $\langle \bar{q}q \rangle=-(0.24\pm 0.01
\,\rm{GeV})^3$, $\langle \bar{s}s \rangle=(0.8\pm 0.2 )\langle
\bar{q}q \rangle$, $\langle \bar{s}g_s\sigma Gs \rangle=m_0^2\langle
\bar{s}s \rangle$, $m_0^2=(0.8 \pm 0.2)\,\rm{GeV}^2$, and $\langle
\frac{\alpha_s GG}{\pi}\rangle=(0.33\,\rm{GeV})^4 $  at the energy
scale $\mu=1\, \rm{GeV}$ \cite{SVZ79,PRT85,Ioffe2005}.

 The  threshold parameters and the Borel parameters
are taken as $s_0=(10.5\pm1.0)\,\rm{GeV}^2$ and
$M^2=(2.2-3.2)\,\rm{GeV}^2$ in the charm channels, and
$s_0=(44.5\pm1.0)\,\rm{GeV}^2$ and $M^2=(5.0-6.0)\,\rm{GeV}^2$ in
the bottom channels, which are determined by the two-point QCD sum
rules  to avoid possible  contaminations from the high resonances
and continuum states \cite{Wang0809Omega,Wang0704}. In
Refs.\cite{MelikhovPLB,MelikhovPRD}, Melikhov et al study the
ground-state form-factor in an exactly solvable harmonic-oscillator
model to illustrate the exact effective continuum threshold for
vacuum-to-hadron correlation function is very difficult to obtain,
as the effective continuum threshold maybe
 depend on the Borel parameter.    We show the values of the strong
coupling constants $g_1$, $g_2$ and $\rm{G}3$ with variation of the
threshold parameters $s_0$  in Fig.1. From the figure we can see
that in the present case  the numerical results are insensitive to
the threshold parameters.

\begin{figure}
 \centering
  \includegraphics[totalheight=5cm,width=6cm]{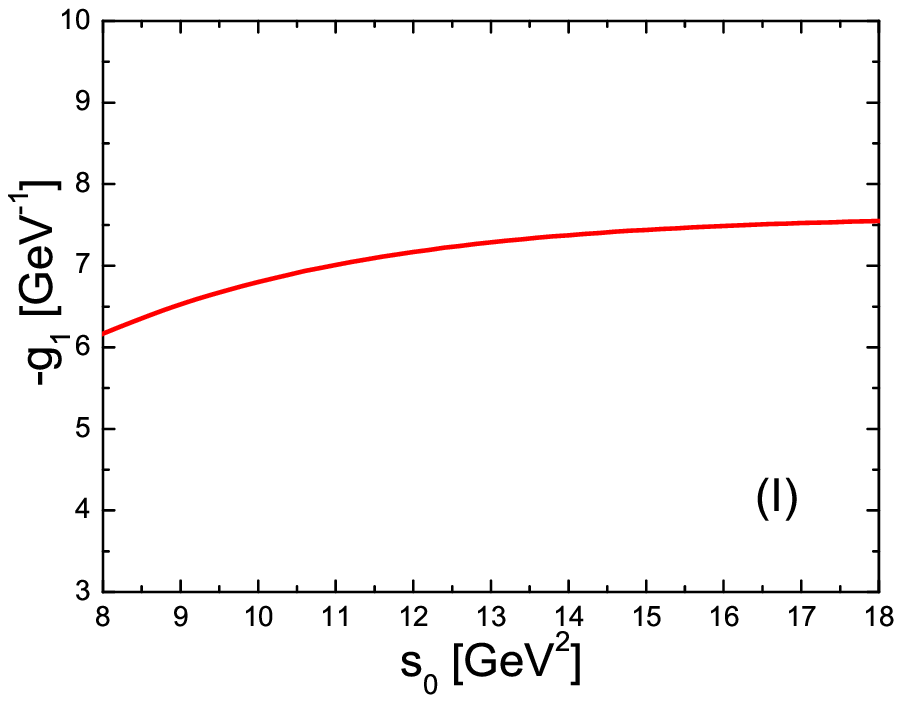}
 \includegraphics[totalheight=5cm,width=6cm]{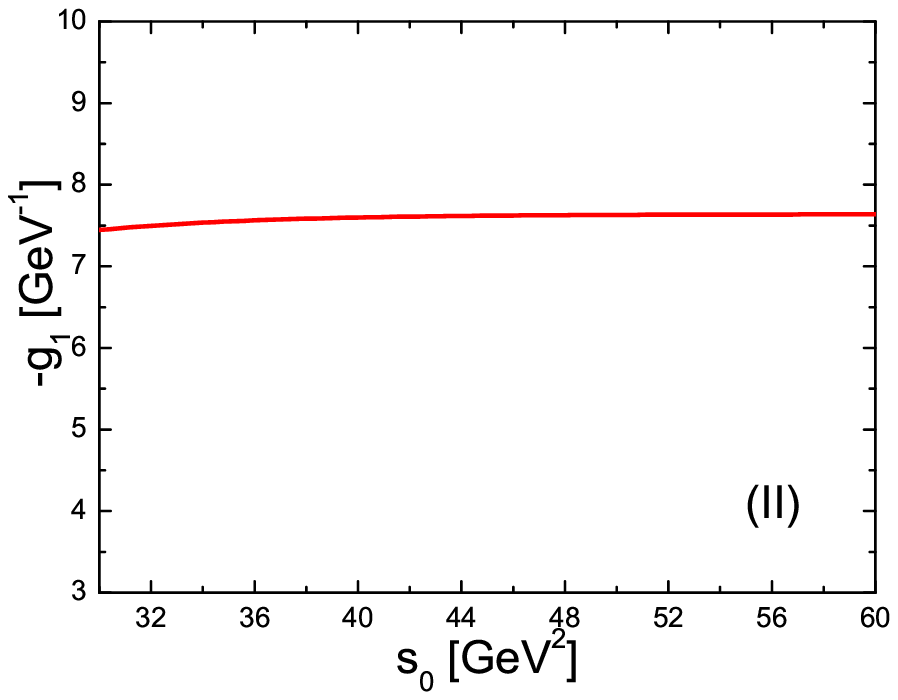}
  \includegraphics[totalheight=5cm,width=6cm]{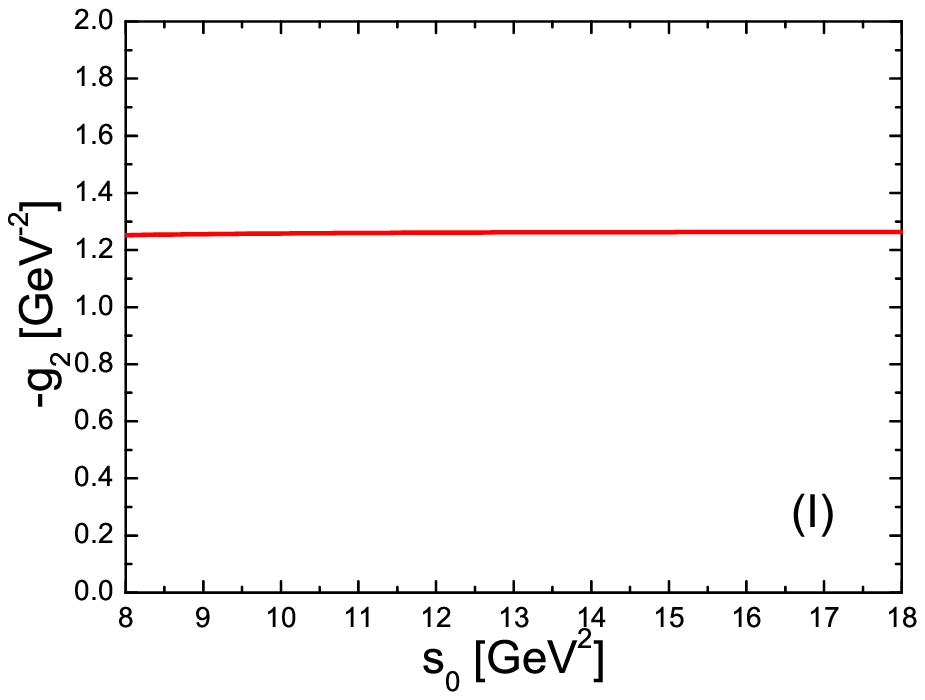}
  \includegraphics[totalheight=5cm,width=6cm]{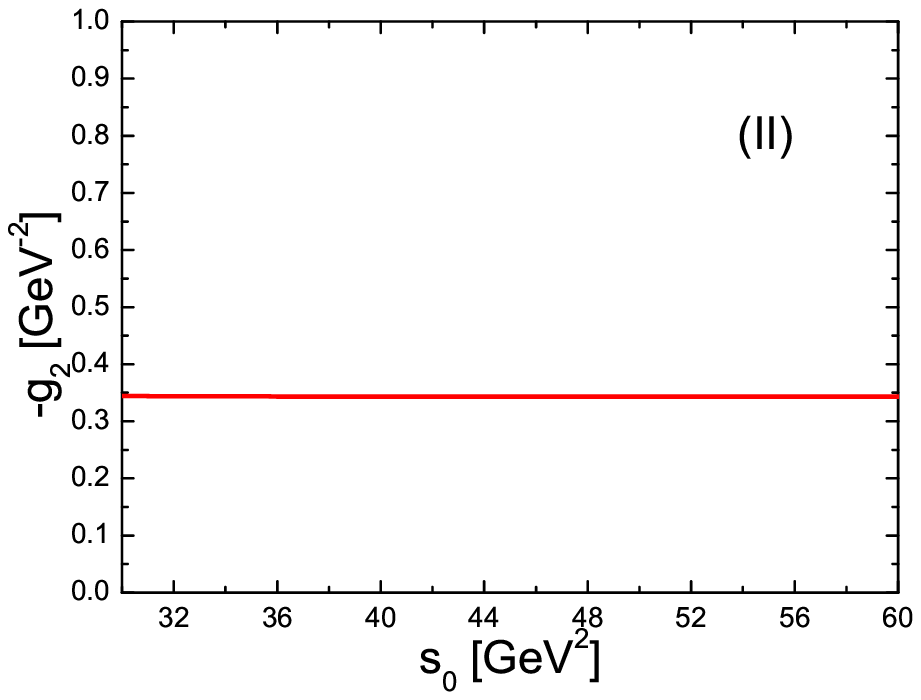}
   \includegraphics[totalheight=5cm,width=6cm]{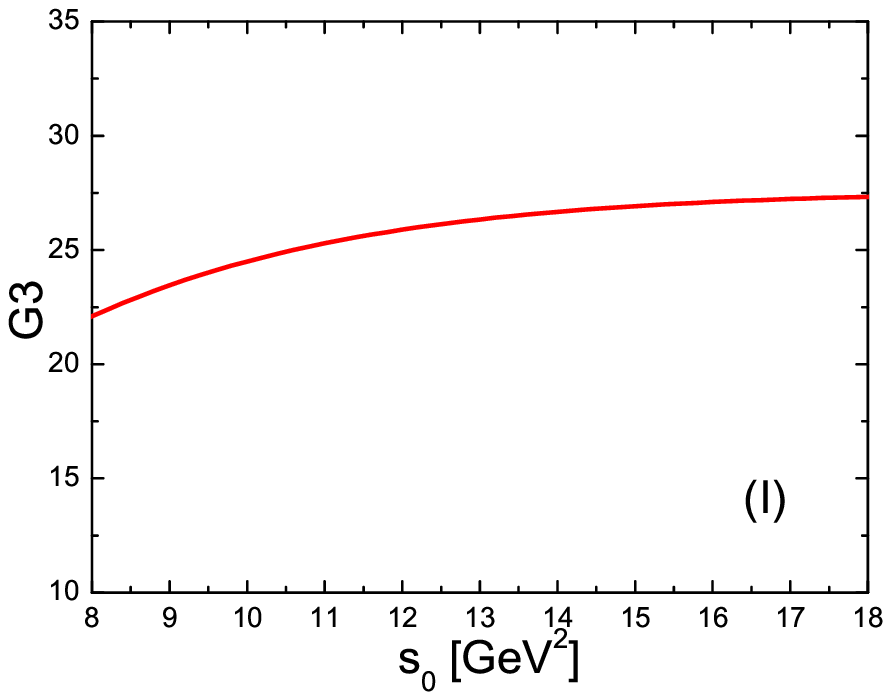}
   \includegraphics[totalheight=5cm,width=6cm]{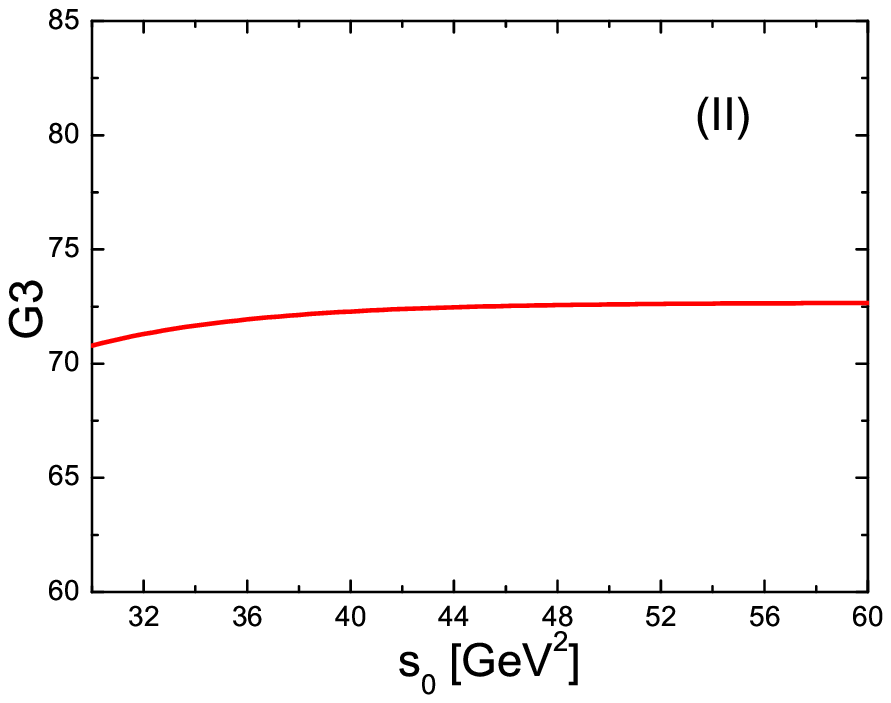}
     \caption{ The strong coupling constants $g_1$, $g_2$ and $\rm{G}3$ with variation of the threshold
     parameters $s_0$, the  Borel parameters are taken to be the central values; the (I) and (II)
     correspond to the charm and bottom channels respectively.  }
\end{figure}

\begin{figure}
 \centering
 \includegraphics[totalheight=5cm,width=6cm]{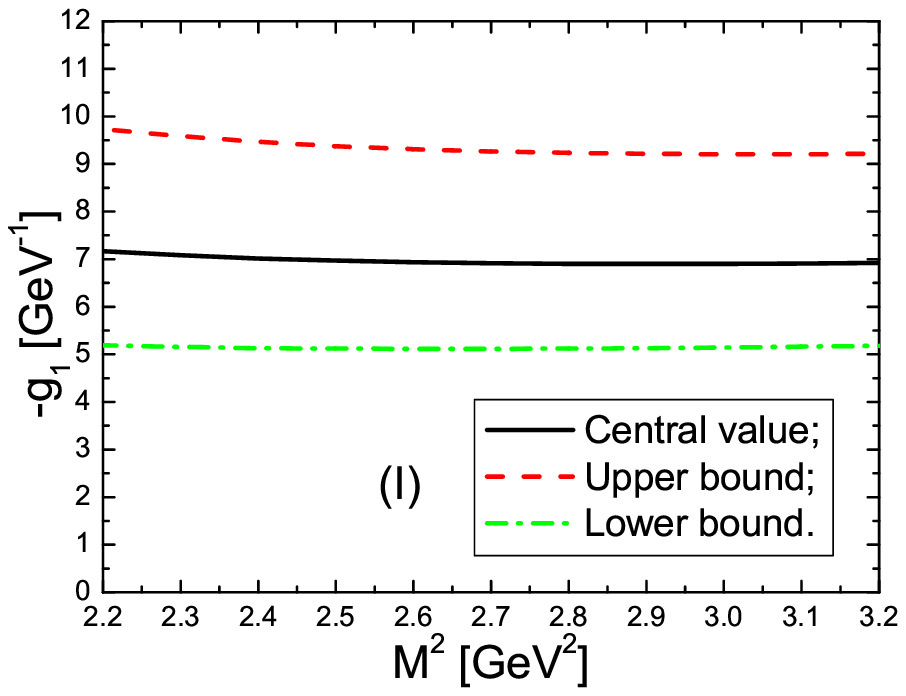}
 \includegraphics[totalheight=5cm,width=6cm]{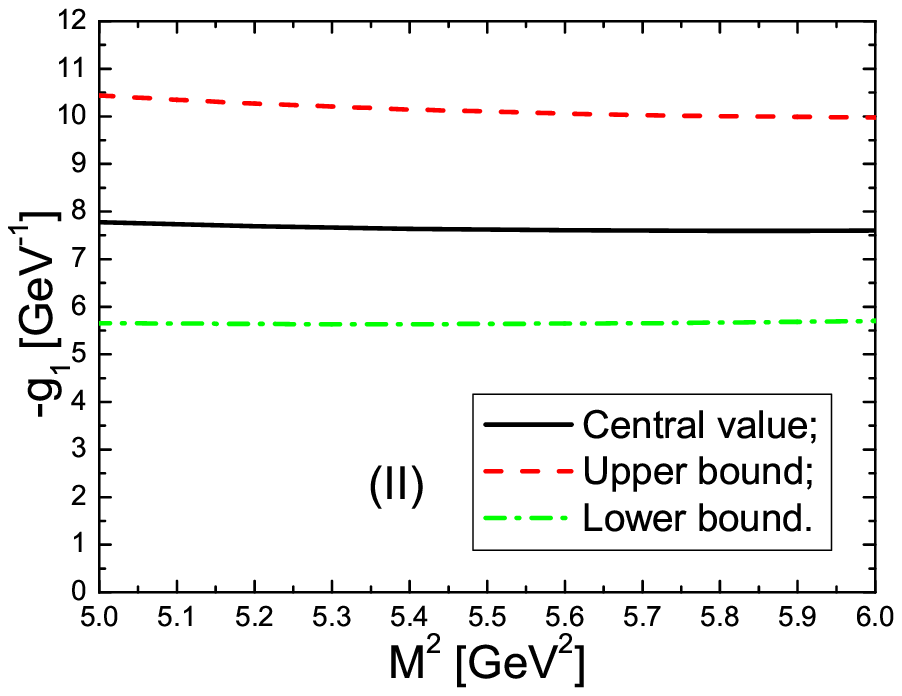}
 \includegraphics[totalheight=5cm,width=6cm]{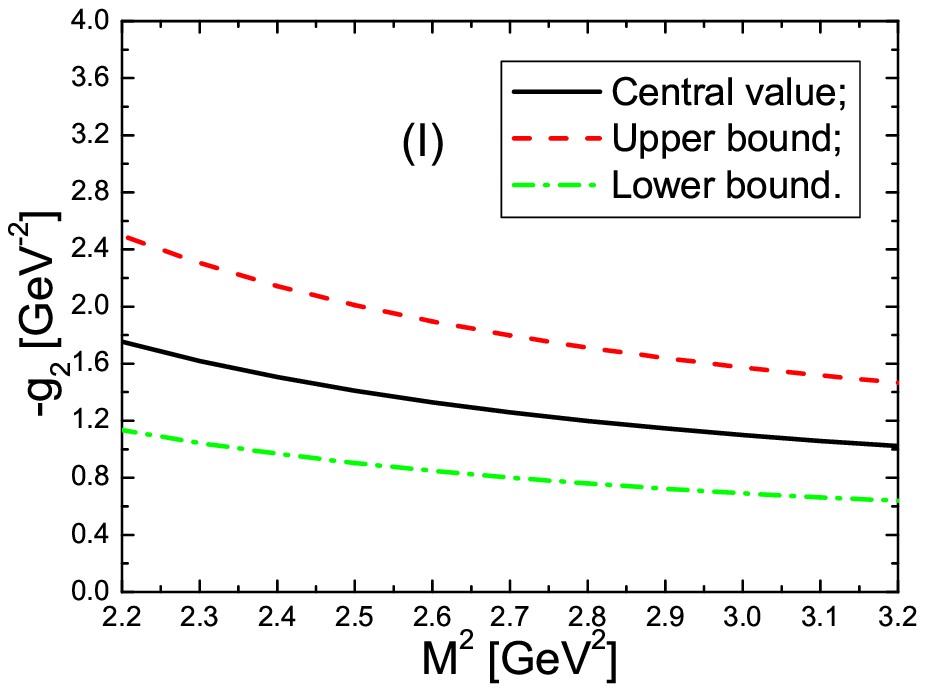}
 \includegraphics[totalheight=5cm,width=6cm]{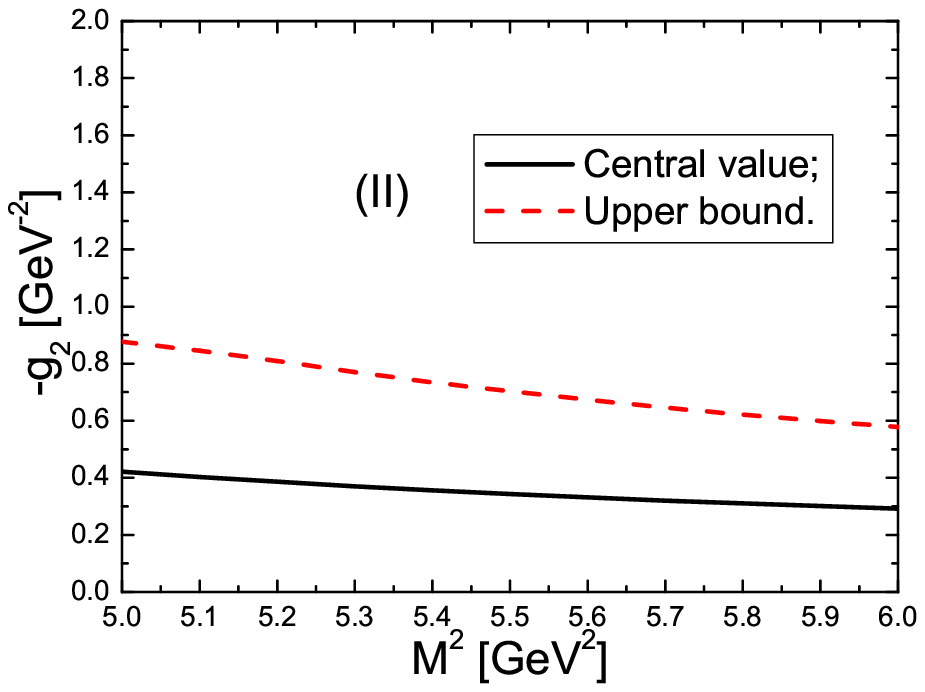}
 \includegraphics[totalheight=5cm,width=6cm]{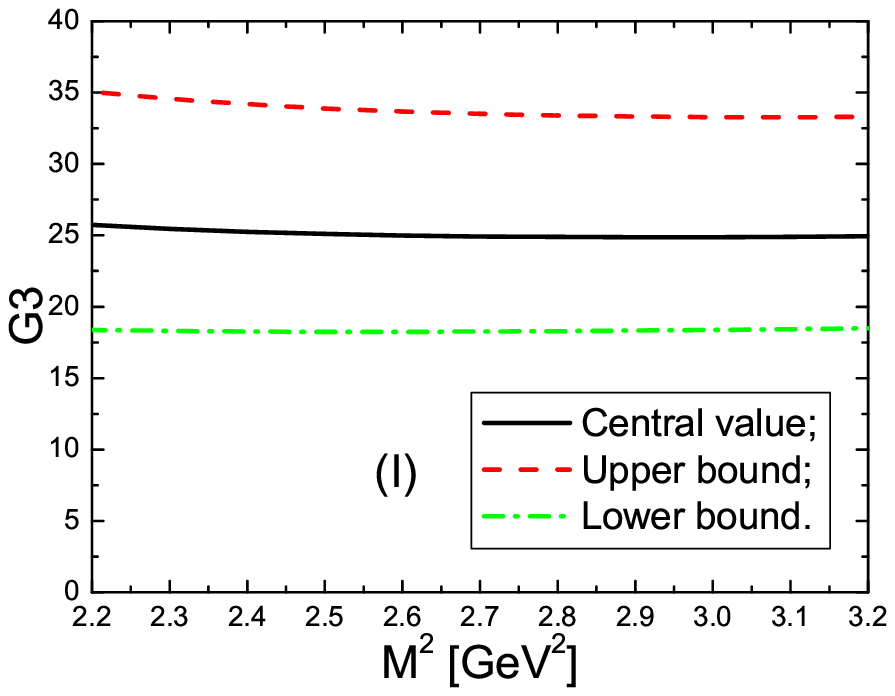}
 \includegraphics[totalheight=5cm,width=6cm]{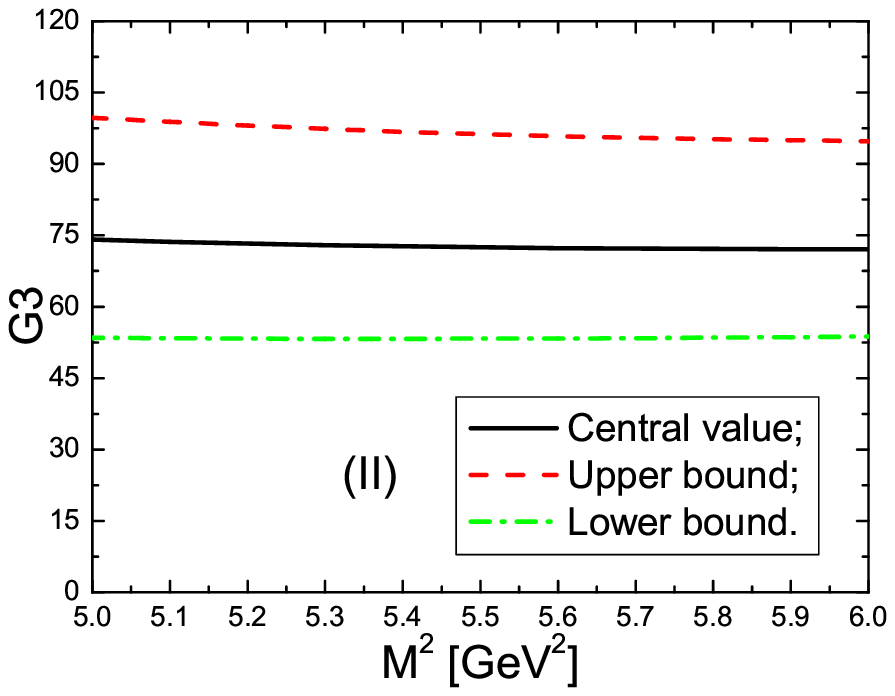}
     \caption{ The strong coupling constants $g_1$, $g_2$ and $\rm{G}3$ with variation of the Borel parameter $M^2$; the (I) and (II)
     correspond to the charm and bottom channels respectively.  }
\end{figure}

 The main uncertainties come
from the six parameters $\lambda_{\Omega_Q}$, $\lambda_{\Omega_Q^*}$
and  $m_{Q}$,
  the variations of those parameters can lead to relatively large
changes for the numerical values,  refining those parameters are of
great importance. Although there are many parameters in the
light-cone distributions amplitudes \cite{VMLC2003,VMLC2007}, the
uncertainties originate from those parameters are rather small.

Taking into account all the uncertainties of the relevant
parameters, finally we obtain the numerical results of the strong
coupling constants $g_1$, $g_2$ and $\rm{G}3$, which are shown in
Fig.2,
\begin{eqnarray}
  -g_1 &=&6.95_{-1.84}^{+2.78} \,\rm{GeV}^{-1} \,  , \nonumber\\
  -g_2 &=&1.35_{-0.71}^{+1.14} \,\rm{GeV}^{-2} \, ,\nonumber \\
\rm{G}3 &=&25.0_{-6.7}^{+10.1}    \, ,
 \end{eqnarray}
 and
\begin{eqnarray}
   -g_1 &=&7.63_{-2.00}^{+2.80} \,\rm{GeV}^{-1} \,  , \nonumber\\
  -g_2 &=&0.34_{-0.34}^{+0.50} \,\rm{GeV}^{-2} \, ,\nonumber \\
\rm{G}3 &=&72.5_{-19.2}^{+27.2}    \, ,
 \end{eqnarray}
 in the charm and bottom channels respectively.
 In this article,  we calculate the uncertainties $\delta$  with the
formula
\begin{eqnarray}
\delta=\sqrt{\sum_i\left(\frac{\partial f}{\partial
x_i}\right)^2\mid_{x_i=\bar{x}_i} (x_i-\bar{x}_i)^2}\,  ,
\end{eqnarray}
 where the $f$ denote  strong coupling constants $g_1$, $g_2$ and
$\rm{G}3$,  the $x_i$ denote the relevant    parameters $m_Q$,
$\langle \bar{q}q \rangle$, $\langle \bar{s}s \rangle$, $\cdots$. As
the partial
 derivatives   $\frac{\partial f}{\partial x_i}$ are difficult to carry
out analytically, we take the  approximation $\left(\frac{\partial
f}{\partial x_i}\right)^2 (x_i-\bar{x}_i)^2\approx
\left[f(\bar{x}_i\pm \Delta x_i)-f(\bar{x}_i)\right]^2$ in the
numerical calculations.

The light-cone  QCD sum rules  approaches have been applied to
determine the strong coupling constant $g_{D^* D \pi}$ in the strong
decay $D^{*+} \to D^0 \pi^+$ both in the leading approximation
\cite{Belyaev94,Colangelo98} and  the next-to-leading order
approximation \cite{Khodjamirian99}. The discrepancy between the
experimental data from the CLEO collaboration and the theoretical
predictions is rather  large.
 The upper bound $g_{D^\ast D \pi}=13.5 $ ($g_{D^\ast D \pi}=10.5 \pm 3.0$  \cite{Khodjamirian99}) is too small to
  account for the experimental data, $g_{D^\ast D \pi}=17.9 \pm 0.3 \pm
1.9$ \cite{CLEO1,CLEO2}.  There  have been several explanations, for
example,    Becirevic et al take into account the contribution from
an explicit radial excitation to the hadronic spectral density to
improve the value of $g_{D^* D \pi}$ \cite{Becirevic03};   Kim tries
to subtract the term $M^2e^{-\frac{s_0}{M^2}}$ which is supposed to
come from a mathematically spurious term and  should not be a part
of the final sum rules to smear the discrepancy \cite{Kim03Thr};
while Duraes et al resort to the intermediate hadronic loops  to
improve the predictive   ability \cite{Duraes04}. Or   the simple
quark-hadron duality ansatz which works in the one-variable
dispersion relation might be too crude for the double dispersion
relation \cite{KhodjamirianConf}. Irrespective of the possible
reasons, the light-cone sum rules cannot give satisfactory value to
account for the experimental data in the channel $D^{*+} \to D^0
\pi^+$, while the light-cone QCD sum rules are rather successful in
calculating the strong coupling constants among the baryons, baryons
and mesons, for example,
 we study the strong decays $\Delta^{++} \to p \pi$,
$\Sigma^*\to \Sigma \pi$ and $\Sigma^*\to \Lambda \pi$ using the
light-cone QCD sum rules,  and observe that the numerical values of
the widths are in agreement with the experimental data within
uncertainties \cite{Wang0707,Wang0809}. The present predictions for
the values of the strong coupling constants $g_1$, $g_2$ and
$\rm{G}3$ are reasonable.

The radiative decays $\Omega_Q^*\to \Omega_Q \gamma$ can be
described by the following electromagnetic lagrangian $\mathcal{L}$,
\begin{eqnarray}
\mathcal{L}&=&-eQ_b \bar{b}\gamma_\mu b A^\mu-eQ_c \bar{c}\gamma_\mu
c A^\mu-eQ_s \bar{s}\gamma_\mu s A^\mu \, ,
\end{eqnarray}
where the $A_\mu$ is the electromagnetic field. From the lagrangian
$\mathcal{L}$,  we can obtain the decay amplitude with the
assumption of the vector meson dominance,
\begin{eqnarray}
\langle \Omega_Q(p)\gamma(q)|\mathcal
{L}|\Omega_Q^*(p+q)\rangle&=&-eQ_s\eta^*_\mu\langle \Omega_Q(p)|
\bar{s}\gamma^\mu s |\Omega_Q^*(p+q)\rangle +\cdots\nonumber \\
&=& -eQ_s\eta^*_\mu f_\phi M_\phi \epsilon_\mu
\frac{i}{q^2-M_\phi^2} \langle\phi(q) \Omega_Q(p)
 |\Omega_Q^*(p+q)\rangle+\cdots \nonumber \\
&=&  \ \frac{ieQ_s\eta_\alpha^* f_\phi }{M_\phi}
U(p)\Gamma^{\alpha\beta}U_\beta(p+q) +\cdots \, ,
\end{eqnarray}
 where the
$\eta_\mu$ is the polarization vector of the photon. In the heavy
quark limit, the matrix elements $ \langle \Omega_Q(p)|{\bar
Q}\gamma_\mu Q|\Omega_Q^*(p+q) \rangle\propto
M_{J/\psi(\Upsilon)}^{-\frac{3}{2}} $ and can be neglected, so we
consider only the contribution of the intermediate $\phi(1020)$.

From the strong coupling constants $g_{1}$ and $g_2$, we can obtain
the decay widths $\Gamma_{\Omega_Q^*\to \Omega_Q \gamma}$,
\begin{eqnarray}
\Gamma_{\Omega_Q^*\to \Omega_Q \gamma}&=&\frac{\alpha
\left(M_{\Omega_Q^*}^2-M_{\Omega_Q}^2\right)}{16 M_{\Omega_Q^*}^3 }
\left(\frac{Q_sf_\phi}{M_\phi}\right)^2 \sum_{ss'} \mid
\eta_\mu^*\overline{U}(p,s) \Gamma^{\mu\nu} U_\nu(p+q,s') \mid^2 \,
,
\end{eqnarray}
the numerical values are
\begin{eqnarray}
\Gamma_{\Omega_c^*\to \Omega_c \gamma}&=&1.16^{+1.12}_{-0.54}\,\rm{KeV} \, ,\nonumber\\
\Gamma_{\Omega_b^*\to \Omega_b
\gamma}&=&0.74^{+0.64}_{-0.34}\,\rm{eV} \, .
\end{eqnarray}
  Here we take the
 value $M_{\Omega_b}=6.0544\,\rm{GeV}$ from the CDF collaboration
 \cite{OmegabCDF}, if we take the value $M_{\Omega_b}=6.615\,\rm{GeV}$ from the D0
 collaboration \cite{OmegabD0}, the radiative decay $\Omega_b^* \to \Omega_b
 \gamma$ is  kinematically  forbidden. Comparing with the values
 from the consitutent quark model $\Gamma_{\Omega_c^*\to \Omega_c
 \gamma}=3.13\,\rm{KeV}$ \cite{CQMwidth}, the hyper central
model $\Gamma_{\Omega_c^*\to \Omega_c
 \gamma}=0.79\,\rm{KeV}$ \cite{CQMwidth}, and the non-relativistic  potential model $\Gamma_{\Omega_c^*\to \Omega_c
 \gamma}=0.36\,\rm{KeV}$ \cite{Dey1994}, the present prediction
  $\Gamma_{\Omega_c^*\to \Omega_c \gamma}=1.16^{+1.12}_{-0.54}\,\rm{KeV}$
  is  rather good, though the uncertainty  is somewhat large.

\section{Conclusion}

In this article, we parameterize the vertexes $\Omega_Q^*\Omega_Q
\phi$ with three tensor structures  due to  Lorentz invariance,
study the corresponding three strong coupling constants with the
light-cone QCD sum rules, then assume the vector meson dominance of
the intermediate $\phi(1020)$ as the contributions from the $J/\psi$
and $\Upsilon$ are  negligible in the heavy quark limit, and
calculate the radiative decay widths $\Gamma_{\Omega_Q^*\to \Omega_Q
\gamma}$. The predictions can be compared with the experimental data
in the future. The strong coupling constants in the vertexes
$\Omega_Q^*\Omega_Q \phi$ are basic parameters in describing the
interactions among the heavy baryon states, once reasonable values
are obtained, we can use them to perform phenomenological analysis.

\section*{Acknowledgment}
This  work is supported by National Natural Science Foundation,
Grant Number 10775051, and Program for New Century Excellent Talents
in University, Grant Number NCET-07-0282, and a foundation of NCEPU.

\section*{Appendix}

The light-cone distribution amplitudes of the $\phi(1020)$ meson are
defined   by \cite{VMLC981,VMLC982,VMLC2003,VMLC2007},
 \begin{eqnarray}
\langle 0| {\bar s} (x) \gamma_\mu s(0) |\phi(q)\rangle& =& q_\mu
 \frac{\epsilon \cdot x}{q \cdot x} f_\phi M_\phi\int_0^1
du  e^{-i \bar{u} q\cdot x}
\left[\phi_{\parallel}(u)+\frac{M_\phi^2x^2}{16}
A(u)\right]\nonumber\\
&&+\left[ \epsilon_\mu-q_\mu \frac{\epsilon \cdot x}{q \cdot x}
\right]f_\phi M_\phi
\int_0^1 du  e^{-i \bar{u}q \cdot x} g_{\perp}^{(v)}(u)  \nonumber\\
&& -\frac{1}{2}x_\mu \frac{\epsilon \cdot x}{(q \cdot x)^2} f_\phi
M_\phi^3 \int_0^1 du e^{-i\bar{u}q \cdot x}C(u)  \, , \nonumber\\
 \langle 0| {\bar s} (x) \sigma_{\mu\nu} s(0) |\phi(q)\rangle& =&
\left[\epsilon_\mu q_\nu-\epsilon_\nu q_\mu\right]i f_\phi^\perp
\int_0^1 du e^{-i \bar{u} q\cdot x}
\left[\phi_{\perp}(u)+\frac{M_\phi^2x^2}{16}
A_\perp(u)\right]\nonumber\\
&&+\left[q_\mu x_\nu-q_\nu x_\mu\right]\frac{\epsilon \cdot x}{(q
\cdot x)^2}i f_\phi^\perp M_\phi^2
\int_0^1 du  e^{-i \bar{u} q \cdot x} B_{\perp}(u)  \nonumber\\
&& + \frac{\epsilon_\mu x_\nu-\epsilon_\nu x_\mu}{2q \cdot x}i
f_\phi^\perp M_\phi^2 \int_0^1 du e^{-i\bar{u}q \cdot x}C_\perp(u)\,
,
\nonumber \\
  \langle 0| {\bar s} (x)\gamma_\mu \gamma_5  s(0) |\phi(q)\rangle  &=& -\frac{1}{
4}\epsilon_{\mu\nu\alpha\beta}\epsilon^\nu q^\alpha
x^\beta\widetilde{f}_\phi M_\phi
  \int_0^1
du e^{-i \bar{u} q \cdot x} g_{\perp}^{(a)}(u) \, , \nonumber\\
  \langle 0| {\bar s} (x)  s(0) |\phi(q)\rangle  &=& -\frac{i}{
2}\epsilon \cdot x\widetilde{f}_\phi^\perp M_\phi^2  \int_0^1 du
e^{-i \bar{u} q \cdot x} h_{\parallel}^{(s)}(u)  \, ,
 \end{eqnarray}
\begin{eqnarray}
\langle 0| {\bar s} (x)\gamma_\alpha \gamma_5
\widetilde{G}_{\mu\nu}(vx) s(0) |\phi(q)\rangle
&=&q_\alpha\left[\epsilon_\mu q_\nu-\epsilon_\nu q_\mu\right] f_\phi
M_\phi \int \mathcal{D}\alpha e^{-i
(\alpha_{\bar{s}}+v\alpha_g) q\cdot x} \mathcal{A}(\alpha_i) \, ,\nonumber\\
\langle 0| {\bar s} (x)i\gamma_\alpha G_{\mu\nu}(vx) s(0)
|\phi(q)\rangle &=&q_\alpha\left[\epsilon_\mu q_\nu-\epsilon_\nu
q_\mu\right] f_\phi M_\phi \int \mathcal{D}\alpha e^{-i
(\alpha_{\bar{s}}+v\alpha_g) q\cdot x} \mathcal{V}(\alpha_i) \, ,\nonumber\\
\langle 0| {\bar s} (x)i\gamma_5 \widetilde{G}_{\mu\nu}(vx) s(0)
|\phi(q)\rangle &=&\left[\epsilon_\mu q_\nu-\epsilon_\nu
q_\mu\right] if_\phi^\perp M_\phi^2 \int \mathcal{D}\alpha e^{-i
(\alpha_{\bar{s}}+v\alpha_g) q\cdot x} \widetilde{\mathcal{S}}(\alpha_i) \, ,\nonumber\\
\langle 0| {\bar s} (x) G_{\mu\nu}(vx) s(0) |\phi(q)\rangle
&=&\left[\epsilon_\mu q_\nu-\epsilon_\nu q_\mu\right] if_\phi^\perp
M_\phi^2 \int \mathcal{D}\alpha e^{-i
(\alpha_{\bar{s}}+v\alpha_g) q\cdot x} \mathcal{S}(\alpha_i) \, ,\nonumber\\
\langle 0| {\bar s} (x)\sigma_{\alpha\beta} G_{\mu\nu}(vx) s(0)
|\phi(q)\rangle &=& \left[ q_\alpha q_\mu g_{\beta\nu}-q_\beta q_\mu
g_{\alpha\nu} -q_\alpha q_\nu
g_{\beta\mu} +q_\beta q_\nu g_{\alpha\mu}\right]   \nonumber\\
&&f_\phi^\perp M_\phi^2 \frac{\epsilon\cdot x}{2 q \cdot x}
\int\mathcal{D}\alpha e^{-i (\alpha_{\bar{s}}+v\alpha_g) q\cdot x}
\mathcal{T}(\alpha_i) \, ,
\end{eqnarray}
where
\begin{eqnarray}
C(u)&=&g_3(u)+\phi_{\parallel}(u)-2g_\perp^{(v)} \nonumber\\
B_\perp(u)&=&h_{\parallel}^{(t)}(u)-\frac{1}{2}\phi_{\perp}(u)-\frac{1}{2}h_3(u) \nonumber\\
C_\perp(u)&=&h_3(u)-\phi_{\perp}(u) \nonumber\\
\int\mathcal{D}\alpha&=& \int_0^1
d\alpha_{\bar{s}}d\alpha_{s}d\alpha_{g}\delta(\alpha_{\bar{s}}+\alpha_{s}+\alpha_{g}-1)
\, ,
\end{eqnarray}
$\bar{u}=1-u$, $\widetilde{f}_\phi=f_\phi-f_\phi^\perp
\frac{2m_s}{M_\phi}$, $\widetilde{f}_\phi^\perp=f_\phi^\perp-f_\phi
\frac{2m_s}{M_\phi}$, the lengthy expressions of the light-cone
distribution amplitudes $\phi_{\parallel}(u)$, $\phi_{\perp}(u)$,
$A(u)$, $A_\perp(u)$, $g_\perp^{(v)}(u)$,
  $g_\perp^{(a)}(u)$, $h_{\parallel}^{(s)}(u)$,
  $h_{\parallel}^{(t)}(u)$, $h_3(u)$, $g_3(u)$, $\mathcal{A}(\alpha_i)$, $\mathcal {S}(\alpha_i)$,
  $\widetilde{\mathcal{S}}(\alpha_i)$,
  $\mathcal{T}(\alpha_i)$, $\mathcal{V}(\alpha_i)$ can be found in
  Refs. \cite{VMLC2003,VMLC2007}.

The six sum rules for the strong coupling constants $g_1$, $g_2$ and
$\rm{G}3$,
\begin{eqnarray}
g_1&=&\frac{1}{\lambda_{\Omega_Q}\lambda_{\Omega_Q^*}
\left(M_{\Omega_Q}+M_{\Omega_Q^*}\right)}\exp{\frac{M_{\Omega_Q}^2+M_{\Omega_Q^*}^2-2u_0(1-u_0)M_\phi^2}{2M^2}}\nonumber \\
&&\left\{-\frac{u_0f_\phi M_\phi
g_{\perp}^{(v)}(1-u_0)}{2\pi^2}M^4E_1(x)\int_0^1 dt t(1-t) e^{-\frac{\widetilde{m}_Q^2}{M^2}}\right.\nonumber\\
 &&+ \frac{u_0m_Q^2f_\phi M_\phi
g_{\perp}^{(v)}(1-u_0)}{36M^2}\langle\frac{\alpha_sGG}{\pi}\rangle\int_0^1 dt \frac{1-t}{t^2} e^{-\frac{\widetilde{m}_Q^2}{M^2}}\nonumber\\
&&-\frac{m_sf_\phi^{\perp}
\phi_{\perp}(1-u_0)}{2\pi^2}M^4E_1(x)\int_0^1 dt t e^{-\frac{\widetilde{m}_Q^2}{M^2}}\nonumber\\
 &&+ \frac{m_sm_Q^2f_\phi^{\perp}
\phi_{\perp}(1-u_0)}{36M^2}\langle\frac{\alpha_sGG}{\pi}\rangle\int_0^1
dt \frac{1}{t^2} e^{-\frac{\widetilde{m}_Q^2}{M^2}} \nonumber \\
&& +\frac{u_0m_sf_\phi^{\perp}M_\phi^2
\widetilde{C}_{\perp}(1-u_0)}{2\pi^2}M^2E_0(x)\int_0^1 dt t e^{-\frac{\widetilde{m}_Q^2}{M^2}}\nonumber\\
 &&- \frac{u_0m_sm_Q^2f_\phi^{\perp}M_\phi^2
\widetilde{C}_{\perp}(1-u_0)}{36M^4}\langle\frac{\alpha_sGG}{\pi}\rangle\int_0^1 dt \frac{1}{t^2} e^{-\frac{\widetilde{m}_Q^2}{M^2}}\nonumber\\
&&-\frac{u_0\widetilde{f}_\phi M_\phi
}{8\pi^2}M^4E_1(x) \frac{d}{du_0}g_\perp^{(a)}(1-u_0)\int_0^1 dt t(1-t) e^{-\frac{\widetilde{m}_Q^2}{M^2}}\nonumber\\
 &&+\frac{u_0m_Q^2\widetilde{f}_\phi M_\phi
}{144M^2}\langle \frac{\alpha_sGG}{\pi}\rangle \frac{d}{du_0}g_\perp^{(a)}(1-u_0)\int_0^1 dt \frac{1-t}{t^2} e^{-\frac{\widetilde{m}_Q^2}{M^2}}\nonumber\\
&&-\frac{\widetilde{f}_\phi M_\phi g_\perp^{(a)}(1-u_0)
}{4\pi^2}M^4E_1(x) \int_0^1 dt t  e^{-\frac{\widetilde{m}_Q^2}{M^2}}\nonumber\\
 &&+\frac{m_Q^2\widetilde{f}_\phi M_\phi
g_\perp^{(a)}(1-u_0)}{72M^2}\langle \frac{\alpha_sGG}{\pi}\rangle  \int_0^1 dt \frac{1}{t^2} e^{-\frac{\widetilde{m}_Q^2}{M^2}}\nonumber\\
&&  +\frac{u_0f_\phi M_\phi^3 }{4\pi^2} M^2E_0(x)\int_0^1 dt
t\int_0^{u_0} d\alpha_{\bar{s}}
\int_{u_0-\alpha_{\bar{s}}}^{1-\alpha_{\bar{s}}}  d\alpha_g \frac{(1-2v)\mathcal{A}(\alpha_i)+\mathcal{V}(\alpha_i)}{\alpha_g}  e^{-\frac{\widetilde{m}_Q^2}{M^2}}   \nonumber\\
&& \left. +\frac{f_\phi M_\phi }{4\pi^2} M^4E_1(x)\int_0^1 dt t
\frac{d}{du_0}\int_0^{u_0} d\alpha_{\bar{s}}
\int_{u_0-\alpha_{\bar{s}}}^{1-\alpha_{\bar{s}}}  d\alpha_g
(1-v)\frac{\mathcal{A}(\alpha_i)+\mathcal{V}(\alpha_i)}{\alpha_g}
e^{-\frac{\widetilde{m}_Q^2}{M^2}}  \right\} \nonumber
 \end{eqnarray}
 \begin{eqnarray}
&+&\frac{1}{\lambda_{\Omega_Q}\lambda_{\Omega_Q^*}
\left(M_{\Omega_Q}+M_{\Omega_Q^*}\right)}\exp{\frac{M_{\Omega_Q}^2+M_{\Omega_Q^*}^2-2m_Q^2-2u_0(1-u_0)M_\phi^2}{2M^2}}\nonumber \\
&&\left\{-\frac{u_0m_s\langle\bar{s}s\rangle f_\phi M_\phi
g_{\perp}^{(v)}(1-u_0)}{3}  -\frac{2\langle\bar{s}s\rangle
f_\phi^{\perp} M_\phi^2
\widetilde{C}_{\perp}(1-u_0)}{3}  \right.\nonumber\\
&&+ \frac{u_0m_s\langle \bar{s}g_s \sigma Gs\rangle f_\phi M_\phi
g_{\perp}^{(v)}(1-u_0)}{18M^2}\left(1+\frac{m_Q^2}{M^2}\right)
\nonumber\\
 &&+\frac{2\langle\bar{s}s\rangle f_\phi^{\perp}
\phi_{\perp}(1-u_0)}{3} M^2E_0(x)- \frac{\langle \bar{s}g_s \sigma
Gs\rangle f_\phi^{\perp}\phi_{\perp}(1-u_0)}{6}\left(1+\frac{m_Q^2}{M^2}\right)\nonumber\\
 &&+ \frac{m_Q^4 f_\phi^{\perp} M_\phi^2 \langle \bar{s}g_s\sigma G s\rangle
A_{\perp}(1-u_0)}{24M^6} + \frac{m_s f_\phi^{\perp} M_\phi^2
A_{\perp}(1-u_0)}{8\pi^2} M^2 E_0(x)   \nonumber\\
&&- \frac{\langle\bar{s}s\rangle f_\phi^{\perp} M_\phi^2
A_{\perp}(1-u_0)}{6} \left(1+\frac{m_Q^2}{M^2}\right)
+\frac{u_0\langle\bar{s}g_s \sigma Gs\rangle f_\phi^{\perp} M_\phi^2
 \widetilde{C}_{\perp}(1-u_0)}{6M^2} \left(1+\frac{m_Q^2}{M^2}\right)\nonumber\\
&&\left.-\frac{m_s\langle \bar{s}s\rangle\widetilde{f}_\phi M_\phi
g_\perp^{(a)}(1-u_0)}{6} \left(1+\frac{m_Q^2}{M^2}\right)
-\frac{u_0m_s\langle \bar{s}s\rangle\widetilde{f}_\phi M_\phi }{12}
\frac{d}{du_0}g_\perp^{(a)}(1-u_0) \right\} \, ,
 \end{eqnarray}

 \begin{eqnarray}
g_2&=&\frac{1}{\lambda_{\Omega_Q}\lambda_{\Omega_Q^*}}\exp{\frac{M_{\Omega_Q}^2+M_{\Omega_Q^*}^2-2u_0(1-u_0)M_\phi^2}{2M^2}}\nonumber \\
&&\left\{\frac{u_0f_\phi M_\phi
 \left[\widetilde{\phi}_{\parallel}(1-u_0)-\widetilde{g}_{\perp}^{(v)}(1-u_0)\right]}{\pi^2}M^2E_0(x)\int_0^1 dt t(1-t) e^{-\frac{\widetilde{m}_Q^2}{M^2}}\right.\nonumber\\
 &&- \frac{u_0m_Q^2 f_\phi M_\phi
\left[\widetilde{\phi}_{\parallel}(1-u_0)-\widetilde{g}_{\perp}^{(v)}(1-u_0)\right]}{18M^4}\langle\frac{\alpha_sGG}{\pi}\rangle\int_0^1 dt \frac{1-t}{t^2} e^{-\frac{\widetilde{m}_Q^2}{M^2}}\nonumber\\
&& -\frac{u_0f_\phi M_\phi^3 \widetilde{A}(1-u_0)}{4\pi^2} \int_0^1
dt t e^{-\frac{\widetilde{m}_Q^2}{M^2}} +\frac{u_0m_Q^2f_\phi
M_\phi^3 \widetilde{A}(1-u_0)}{72M^6} \langle \frac{\alpha_sGG}{\pi}
\rangle\int_0^1 dt \frac{1}{t^2} e^{-\frac{\widetilde{m}_Q^2}{M^2}} \nonumber\\
 &&-\frac{2u_0m_sf_\phi^{\perp}M_\phi^2
\widetilde{\widetilde{B}}_{\perp}(1-u_0)}{\pi^2}\int_0^1 dt t e^{-\frac{\widetilde{m}_Q^2}{M^2}}\nonumber\\
 &&+ \frac{u_0m_sm_Q^2f_\phi^{\perp}M_\phi^2
\widetilde{\widetilde{B}}_{\perp}(1-u_0)}{9M^6}\langle\frac{\alpha_sGG}{\pi}\rangle\int_0^1 dt \frac{1}{t^2} e^{-\frac{\widetilde{m}_Q^2}{M^2}}\nonumber\\
&&-\frac{u_0\widetilde{f}_\phi M_\phi
}{4\pi^2}M^2E_0(x) g_\perp^{(a)}(1-u_0)\int_0^1 dt t(1-t) e^{-\frac{\widetilde{m}_Q^2}{M^2}}\nonumber\\
 &&+\frac{u_0m_Q^2\widetilde{f}_\phi M_\phi
}{72M^4}\langle \frac{\alpha_sGG}{\pi}\rangle g_\perp^{(a)}(1-u_0)\int_0^1 dt \frac{1-t}{t^2} e^{-\frac{\widetilde{m}_Q^2}{M^2}}\nonumber\\
&&  \left.+\frac{f_\phi M_\phi}{2\pi^2} M^2E_0(x) \int_0^1 dt
t\int_0^{u_0} d\alpha_{\bar{s}}
\int_{u_0-\alpha_{\bar{s}}}^{1-\alpha_{\bar{s}}} d\alpha_g
\frac{\mathcal{A}(\alpha_i)+(1-2v)\mathcal{V}(\alpha_i)} {\alpha_g}
e^{-\frac{\widetilde{m}_Q^2}{M^2}}    \right\} \nonumber
 \end{eqnarray}
 \begin{eqnarray}
&+&\frac{1}{\lambda_{\Omega_Q}\lambda_{\Omega_Q^*}
}\exp{\frac{M_{\Omega_Q}^2+M_{\Omega_Q^*}^2-2m_Q^2-2u_0(1-u_0)M_\phi^2}{2M^2}}\nonumber \\
&&\left\{\frac{2u_0m_s\langle\bar{s}s\rangle f_\phi M_\phi
\left[\widetilde{\phi}_\parallel(1-u_0)-\widetilde{g}_{\perp}^{(v)}(1-u_0)\right]}{3M^2}
  \right.\nonumber\\
&&- \frac{u_0m_s\langle \bar{s} s\rangle f_\phi M_\phi^3
\widetilde{A}(1-u_0)}{6M^4}\left(1+\frac{m_Q^2}{M^2}\right)
\nonumber\\
&&- \frac{u_0m_s\langle \bar{s}g_s \sigma Gs\rangle f_\phi M_\phi
\left[\widetilde{\phi}_\parallel(1-u_0)-\widetilde{g}_{\perp}^{(v)}(1-u_0)\right]}{9M^4}\left(1+\frac{m_Q^2}{M^2}\right)
\nonumber\\
 &&+\frac{8u_0\langle\bar{s}s\rangle f_\phi^{\perp}M_\phi^2
\widetilde{\widetilde{B}}_{\perp}(1-u_0) }{3M^2} - \frac{2u_0\langle
\bar{s}g_s \sigma
Gs\rangle f_\phi^{\perp}M_\phi^2\widetilde{\widetilde{B}}_{\perp}(1-u_0)}{3M^4}\left(1+\frac{m_Q^2}{M^2}\right)\nonumber\\
 &&\left.-\frac{u_0m_s\langle
\bar{s}s\rangle\widetilde{f}_\phi M_\phi g_\perp^{(a)}(1-u_0)}{6M^2}
\right\} \, ,
 \end{eqnarray}

\begin{eqnarray}
\rm{G}3&=&\frac{1}{\lambda_{\Omega_Q}\lambda_{\Omega_Q^*}}\exp{\frac{M_{\Omega_Q}^2+M_{\Omega_Q^*}^2-2u_0(1-u_0)M_\phi^2}{2M^2}}\nonumber \\
&&\left\{\frac{f_\phi M_\phi
 \left[\widetilde{\phi}_{\parallel}(1-u_0)-\widetilde{g}_{\perp}^{(v)}(1-u_0)\right]}{2\pi^2}M^4E_1(x)\int_0^1 dt t(1-t) e^{-\frac{\widetilde{m}_Q^2}{M^2}}\right.\nonumber\\
 &&- \frac{m_Q^2f_\phi M_\phi
\left[\widetilde{\phi}_{\parallel}(1-u_0)-\widetilde{g}_{\perp}^{(v)}(1-u_0)\right]}{36M^2}\langle\frac{\alpha_sGG}{\pi}\rangle\int_0^1 dt \frac{1-t}{t^2} e^{-\frac{\widetilde{m}_Q^2}{M^2}}\nonumber\\
&& -\frac{f_\phi M_\phi^3  \widetilde{A}(1-u_0)}{8\pi^2}
M^2E_0(x)\int_0^1 dt t
e^{-\frac{\widetilde{m}_Q^2}{M^2}}\nonumber\\
 &&+\frac{m_Q^2f_\phi M_\phi^3 \widetilde{A}(1-u_0)}{144M^4}
\langle \frac{\alpha_sGG}{\pi}
\rangle\int_0^1 dt \frac{1}{t^2} e^{-\frac{\widetilde{m}_Q^2}{M^2}} \nonumber\\
&&+\frac{m_sf_\phi^{\perp}
\phi_{\perp}(1-u_0)}{2\pi^2}M^4E_1(x)\int_0^1 dt t
e^{-\frac{\widetilde{m}_Q^2}{M^2}}\nonumber\\
 &&- \frac{m_sm_Q^2f_\phi^{\perp}
\phi_{\perp}(1-u_0)}{36M^2}\langle\frac{\alpha_sGG}{\pi}\rangle\int_0^1 dt \frac{1}{t^2} e^{-\frac{\widetilde{m}_Q^2}{M^2}}\nonumber\\
 &&-\frac{m_sf_\phi^{\perp}M_\phi^2\widetilde{\widetilde{B}}_\perp
 (1-u_0)}{\pi^2}M^2E_0(x)\int_0^1 dt t e^{-\frac{\widetilde{m}_Q^2}{M^2}}\nonumber\\
 &&+ \frac{m_sm_Q^2f_\phi^{\perp}M_\phi^2
\widetilde{\widetilde{B}}_{\perp}(1-u_0)}{18M^4}\langle\frac{\alpha_sGG}{\pi}\rangle\int_0^1 dt \frac{1}{t^2} e^{-\frac{\widetilde{m}_Q^2}{M^2}}\nonumber\\
&&+\frac{\widetilde{f}_\phi M_\phi
}{8\pi^2}M^4E_1(x) g_\perp^{(a)}(1-u_0)\int_0^1 dt t(1+t) e^{-\frac{\widetilde{m}_Q^2}{M^2}}\nonumber\\
 &&-\frac{m_Q^2\widetilde{f}_\phi M_\phi
}{144M^2}\langle \frac{\alpha_sGG}{\pi}\rangle g_\perp^{(a)}(1-u_0)\int_0^1 dt \frac{1+t}{t^2} e^{-\frac{\widetilde{m}_Q^2}{M^2}}\nonumber\\
&&  \left.-\frac{u_0f_\phi M_\phi^3 }{2\pi^2}M^2E_0(x) \int_0^1 dt
t\int_0^{u_0} d\alpha_{\bar{s}}
\int_{u_0-\alpha_{\bar{s}}}^{1-\alpha_{\bar{s}}} d\alpha_g
v\frac{\mathcal{A}(\alpha_i)-\mathcal{V}(\alpha_i)} {\alpha_g}
e^{-\frac{\widetilde{m}_Q^2}{M^2}}    \right\} \nonumber
 \end{eqnarray}
\begin{eqnarray}
&+&\frac{1}{\lambda_{\Omega_Q}\lambda_{\Omega_Q^*}
}\exp{\frac{M_{\Omega_Q}^2+M_{\Omega_Q^*}^2-2m_Q^2-2u_0(1-u_0)M_\phi^2}{2M^2}}\nonumber \\
&&\left\{\frac{m_s\langle\bar{s}s\rangle f_\phi M_\phi
\left[\widetilde{\phi}_\parallel(1-u_0)-\widetilde{g}_{\perp}^{(v)}(1-u_0)\right]}{3}
  \right.\nonumber\\
&&- \frac{m_s\langle \bar{s}s\rangle f_\phi M_\phi^3
\widetilde{A}(1-u_0)}{12M^2}\left(1+\frac{m_Q^2}{M^2}\right)
\nonumber\\
&&- \frac{m_s\langle \bar{s}g_s \sigma Gs\rangle f_\phi M_\phi
\left[\widetilde{\phi}_\parallel(1-u_0)-\widetilde{g}_{\perp}^{(v)}(1-u_0)\right]}{18M^2}\left(1+\frac{m_Q^2}{M^2}\right)
\nonumber\\
 &&-\frac{2\langle\bar{s}s\rangle f_\phi^{\perp}
\phi_{\perp}(1-u_0)}{3}M^2E_0(x) + \frac{\langle \bar{s}g_s \sigma
Gs\rangle f_\phi^{\perp}\phi_{\perp}(1-u_0)}{6}\left(1+\frac{m_Q^2}{M^2}\right)\nonumber\\
 &&- \frac{m_Q^4 f_\phi^{\perp} M_\phi^2 \langle \bar{s}g_s\sigma G s\rangle
A_{\perp}(1-u_0)}{24M^6} - \frac{m_s f_\phi^{\perp} M_\phi^2
A_{\perp}(1-u_0)}{8\pi^2} M^2 E_0(x)   \nonumber\\
&&+ \frac{\langle\bar{s}s\rangle f_\phi^{\perp} M_\phi^2
A_{\perp}(1-u_0)}{6} \left(1+\frac{m_Q^2}{M^2}\right)
+\frac{4\langle\bar{s}s\rangle f_\phi^{\perp}M_\phi^2
\widetilde{\widetilde{B}}_{\perp}(1-u_0) }{3}\nonumber\\
 && - \frac{\langle
\bar{s}g_s \sigma
Gs\rangle f_\phi^{\perp}M_\phi^2\widetilde{\widetilde{B}}_{\perp}(1-u_0)}{3M^2}\left(1+\frac{m_Q^2}{M^2}\right)\nonumber\\
 &&\left.+\frac{m_s\langle
\bar{s}s\rangle\widetilde{f}_\phi M_\phi
g_\perp^{(a)}(1-u_0)}{12}\left(1+\frac{2m_Q^2}{M^2}\right) \right\}
\, ,
 \end{eqnarray}
where $M_1^2=M_2^2=2M^2$ and
$u_0=\frac{M_1^2}{M_1^2+M_2^2}=\frac{1}{2}$ as
$\frac{M_{\Omega_Q^*}^2}{M_{\Omega_Q^*}^2+M_{\Omega_Q}^2}\approx
\frac{1}{2}$,  $v=\frac{u_0-\alpha_{\bar{s}}}{\alpha_g}$,
$\widetilde{m}_Q^2=\frac{m_Q^2}{t}$,
$E_n(x)=1-(1+x+\frac{x^2}{2!}+\cdots+\frac{x^n}{n!})e^{-x}$,
$x=\frac{s_0}{M^2}$, and
$\widetilde{\widetilde{f}}(1-u_0)=\int_0^{u_0}du\int_0^u dt f(1-t)$,
$\widetilde{f}(1-u_0)=\int_0^{u_0}du f(1-u)$, the $f(u)$ denote the
light-cone distribution amplitudes. For some technical details
concerning the three particle $\phi$-meson light-cone distribution
amplitudes, one can consult Ref.\cite{WangJPG}.

\end{document}